# Magnetic field induced phenomena in UIrGe in fields applied along *b* axis

Jiří Pospíšil[1,2], Yoshinori Haga[1], Yoshimitsu Kohama[3], Atsushi Miyake[3], Shinsaku Kambe[1], Naoyuki Tateiwa[1], Michal Vališka[2], Petr Proschek[2], Jan Prokleška[2], Vladimír Sechovský[2], Masashi Tokunaga[3], Koichi Kindo[3], Akira Matsuo[3], and Etsuji Yamamoto[1]

[1]*Advanced Science Research Center, Japan Atomic Energy Agency, Tokai, Ibaraki, 319-1195, Japan*
[2]*Charles University, Faculty of Mathematics and Physics, Department of Condensed Matter Physics, Ke Karlovu 5, 121 16 Prague 2, Czech Republic*
[3]*The Institute for Solid State Physics, University of Tokyo, Kashiwa, Chiba 277-8581, Japan*

**Abstract**

The metamagnetic transition between the antiferromagnetic and paramagnetic state in UIrGe has been studied at various temperatures by magnetization, heat capacity and magnetocaloric-effect measurements on a single crystal in static and pulsed magnetic fields applied along the orthorhombic *b*-axis. A first-order transition is observed at temperatures below 13 K and a second-order one at higher temperatures up to the Néel temperature ($T_N$ = 16.5 K). The first-order transition is accompanied by a dramatic increase of the Sommerfeld coefficient. Magnetization measurements extended to the paramagnetic range revealed an anomalous S-shape (inflection point at a magnetic field $H_m$) in magnetization isotherms at temperatures above 13 K and a temperature dependence of susceptibility with a maximum at $T_{max}$ well above $T_N$. The lines representing the temperature-induced evolution of $H_m$ and field-induced evolution of $T_{max}$, respectively, are bound for the point in the magnetic phase diagram at which the order of metamagnetic transition changes. A tentative scenario explaining these anomalies by antiferromagnetic correlations or short-range order in the paramagnetic state is discussed.

I.  INTRODUCTION

In recent years, the U$T$Ge ($T$ = late transition metal) compounds, which crystallize in the orthorhombic TiNiSi-type structure, have attracted a new wave of interest mainly due to ambient-pressure superconductivity observed in the itinerant 5$f$-electron ferromagnets URhGe and UCoGe[1, 2]. Much less attention has been paid to the isostructural and isoelectronic compound UIrGe, which has an antiferromagnetic (AFM) ground state (Néel temperature $T_N$ = 16.5 K)[3, 4] with a low Sommerfeld coefficient $\gamma$ = 16 mJ/mol K$^2$ similar to other U$T$Ge antiferromagnets[5]. The $\gamma$-values of the two ferromagnets, UCoGe and URhGe are considerably higher, 65 and 100 mJ/mol K$^2$, respectively[6].

A common feature of the U$T$Ge compounds crystallizing in the orthorhombic TiNiSi-type structure are the zig-zag chains of the U nearest-neighbor ions which meander along the $a$ axis. The strong magnetocrystalline anisotropy in these materials is characterized by a hard magnetization direction identified with this axis as a consequence of the 5$f$-electron orbital moment orienting perpendicular to the 5$f$ charge density concentrated within the chain[6].

This is projected also to the anisotropic susceptibility in the paramagnetic (PM) state showing large signals along the $b$ and $c$ axis, respectively and a very low weakly temperature dependent $a$-axis susceptibility[4].

The magnetocrystalline anisotropy in UIrGe at temperatures near the PM ↔ AFM transition is well illustrated by different evolution of a $T_N$-related heat-capacity anomaly in magnetic fields applied along each crystallographic axis. $T_N$ shifts toward lower temperatures with increasing fields applied along $b$ or c. The almost negligible influence of the field applied along $a$ corroborates that the $a$ axis is the hard magnetization direction[7].

At temperatures below $T_N$, metamagnetic transitions (MT) have been observed at critical fields $\mu_0 H_c$ = 21 T and 14 T (at 2 K) applied along $b$ and $c$, respectively[8]. The $a$ axis as the hard magnetization direction is evidenced also by a weak linear magnetization response to magnetic fields up to 51 T[8]. Although $H_c$ is much higher for the $b$ axis then for $c$, the considerably higher slopes of the $b$ axis magnetization curve below and above the metamagnetic transition and the highest magnetic moment measured in 51 T (0.87 $\mu_B$/f.u.) would qualify $b$ as an easy magnetization direction in the AFM state.

A commensurate, non-collinear antiferromagnetic structure with reduced U magnetic moments of 0.36 $\mu_B$/f.u. confined within the $a$-$c$ plane has been reported for UIrGe at temperatures below $T_N$[9]. The AFM structure collapses to a ferromagnetic-like (polarized paramagnetic) ordering of U moments along the $c$ axis in a magnetic field of 14.5 T (> $\mu_0 H_c$) applied along this direction[10].

UIrGe exhibits in fields applied along the $b$ axis some peculiarities which deserve further investigation. The magnetization in the $b$-axis fields higher than $H_c$ reaches considerably higher values than in the corresponding $c$-axis fields[8, 11] although a ferromagnetic-like state in the $c$-axis field above $H_c$ has been suggested from the aforementioned neutron experiment[10].

Heat-capacity data measured in $b$-axis fields up to 17.5 T (Fig. 2 in Ref. 8) show clear evidences of a change of the phase transition at $T_N$ from a second-order type magnetic phase transition (SOMPT) in fields up to 14 T to first-order type (FOMPT) in higher fields.

The temperature dependence of the $b$-axis susceptibility, $\chi_b(T)$, shows a broad maximum at $T_{max}$ = 29 K in contrary to $\chi_a(T)$ and $\chi_c(T)$, which do not exhibit such anomalies well above $T_N$[4].

These interesting findings motivated us to perform detailed measurements of magnetization (susceptibility) in wide ranges of temperatures and $b$-axis magnetic fields to explore the overall evolution of magnetic phase transitions and anomalous behavior in paramagnetic state. Results of magnetization measurements in static fields up to 18.5 T have been combined with reliable magnetization data obtained in pulsed magnetic fields up to 35 T together with a field dependence of heat capacity up to 25 T at 1.6 K. We also measured the magnetocaloric effect (MCE) in pulsed fields at some representative temperatures. MCE data are very useful for assessing reliability of magnetization data measured in pulsed fields, as well as for the resolution between FOMPT and SOMPT.

The results obtained enabled us to create a *T-H* magnetic phase diagram containing important UIrGe properties in *b*-axis magnetic fields including specific regimes in the paramagnetic state.

We have observed a FOMPT (AFM ↔ PM field-induced transition) at low temperatures, which is accompanied by a dramatic increase of Sommerfeld coefficient. This may indicate a Fermi surface reconstruction due to changes of arrangement of itinerant 5*f*-electron magnetic moments. The AFM ↔ PM field-induced transition at temperatures above 13 K has attributes of SOMPT. Near to 13 K a tricritical point (TCP) separating the FOMPT and SOMPT segments in the magnetic phase diagram is expected. We have also found that the $\chi_b(T)$ dependence is anomalous up to considerably higher temperatures than $T_{max}$. It progressively deviates downwards from the modified Curie-Weiss (MCW) law with decreasing temperature below a characteristic $T_{dev}$, which is slightly less than 50 K. Besides the anomalous $\chi_b(T)$ below $T_{dev}$ we have observed in the paramagnetic state also a broad S-shape anomaly on magnetization isotherms *M(H)* at temperatures above TCP. These *M(H)* curves have an inflection point at a characteristic field marked as $H_m$. Interestingly, both dependences, $T_{max}(H)$ with increasing *H* and $H_m(T)$ with decreasing *T*, are heading towards TCP. We propose a tentative scenario which could explain these unusual $\chi_b(T)$ and *M(H)* behaviors in the paramagnetic state by the influence of AFM correlations or short-range AFM order. The paramagnetic phase space, where this mechanism is effective, is demarcated by the SOMPT phase transition (AFM ↔ PM) line, and $T_{dev}(H)$ and $H_m(T)$ crossover lines and we call it Correlated Paramagnetic (CPM) regime in analogy to the notation from Refs.12, 13 in which various aspects of the phase space of AFM correlated electron systems has been thoroughly discussed. We have introduced a Polarized Paramagnetic (PPM) regime in the paramagnetic phase space in fields above $H_c$ of FOMPTs. The $H_m(T)$ line is then within the crossover between CPM and PPM regimes.

## II. EXPERIMENTAL

A single crystal of RRR = 8 along the *b* axis, grown by Czochralski method in a tri-arc furnace from constituent elements (purity of Ir 3N5 and Ge 6N, U-SSE treated), was used for the static field experiments. The batch of UIrGe single crystals grown for the recent high-pressure experiements[14] was used for pulsed-field experiments. The magnetization was measured in steady magnetic fields up to 14 T in a PPMS 14T (Quantum Design) using the VSM option and up to 18.5 T in a cryomagnet (Cryogenic Limited system) equipped with miniature Hall probes (Arepoc Company) sensitive to the dipole field created by the sample magnetization. In order to obtain absolute values, the Hall probe data were scaled to the PPMS 14T data in the overlapping low-field region. The static magnetic susceptibility $\chi$ was calculated from magnetization *M*, and applied-magnetic-field data as $\chi = M/\mu_0 H$.

The magnetization measurements in pulsed magnetic fields were performed using a non-destructive short-pulse magnet with a typical pulse duration of ~36 ms installed at the International Mega Gauss Science Laboratory (IMGSL) of the Institute for Solid State Physics at the University of Tokyo. The magnetization was measured by a conventional induction method using coaxial pick-up coils in fields up to ~35 T. The sample was placed in He gas (at 4.2 K in He liquid).

The magnetocaloric effect (MCE) was measured as a spontaneous temperature change during the magnetic field sweeps in quasi-adiabatic conditions up to a maximum field of ~25 T by using a long-pulse magnet. To minimize the eddy current heating during the magnetic field sweep, the temperature of the sample was measured with a slow field sweep rate of ~40 T/s[15, 16].

The field dependence of the heat capacity was carried out in the identical long-pulse magnet as in previous MCE experiment at nearly isothermal conditions (±0.2 K) using the AC technique[16].

The long-pulse magnet at IMGSL can also produce highly stable flat-top pulsed magnetic fields (±50 Oe) over 100 ms timescale on the top of the field pulse when employing a field-feedback controller[17]. We utilized the flattop region for measuring the temperature dependence of heat capacity by applying the heat-pulse method[18]. The magnetic field was stabilized for ~150 ms at the maximum field and simultaneously1-ms heat pulses were applied to the sample leading to sudden $\Delta T$ jumps of the sample temperature. The heat-capacity values have been obtained from the ratio of corresponding $\Delta Q$ and $\Delta T$ increments, where $\Delta Q = \int dPdt$ and $P$ is the power applied within the heat pulse[17].

All measurements involving application of a magnetic field were performed exclusively with the field direction along the $b$ axis.

## III. EXPERIMENTAL RESULTS

### A. Magnetization

The pronounced anisotropy of magnetic susceptibility in UIrGe is well documented in Fig. 1a. Apparently, the only common feature of all three susceptibility components, $\chi_a(T)$, $\chi_b(T)$, $\chi_c(T)$, each measured in the magnetic field applied along the corresponding crystallographic axis, $a$, $b$, $c$, is a well-defined sharp drop at $T_N = 16.5$ K. This temperature is in agreement with the Néel temperature detected by specific-heat measurements reported in previous works[4, 14, 19]. By far the weakest signal is observed in $H$ applied along the $a$ axis, which represents the hard magnetization direction. The $\chi_b(T)$ and $\chi_c(T)$ values are much higher but their low-temperature dependences differ considerably. $\chi_c(T)$ increases with decreasing temperature following the corresponding MCW fit nearly to $T_N$, similar to $\chi_a(T)$. $\chi_b(T)$ starts to deviate from the MCW fit already at temperature $T_{dev}$, which is around 55 K, and exhibits a broad maximum at $T_{max} = 29$ K. Both, $T_N$ and $T_{max}$, decrease with increasing the magnetic field as can be seen in Fig. 1b.

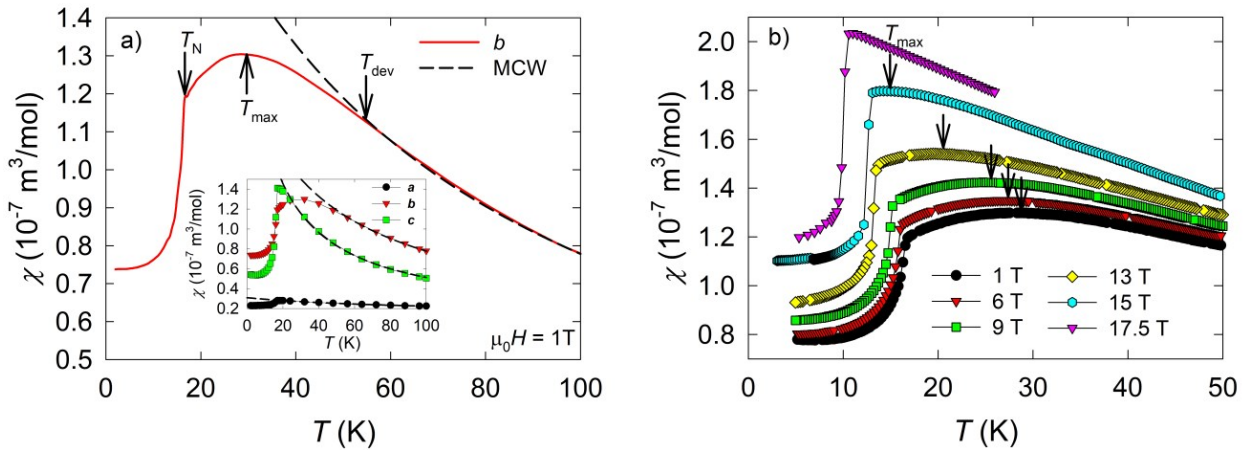

Fig. 1 a) Temperature dependence of the static magnetic susceptibility $\chi$ in a magnetic field of 1 T applied along the $b$ axis. The dashed line represents the fit of data to a MCW law above 50 K.

Inset: Comparison of susceptibilities measured along each of the crystallographic axis. Data have been taken from Ref.4. The dashed lines represent the fits of data to MCW law above 50 K.

b) The $\chi(T)$ dependences in various magnetic fields applied along the $b$ axis. For sake of clarity only selected $\chi(T)$ curves are displayed. The successive curves are mutually shifted upwards by $5 \cdot 10^{-8}$ m$^3$/mol.

The parts of the magnetization isotherms measured in static magnetic fields in intervals, including the anomalies connected with MT between the antiferromagnetic and paramagnetic phase, are plotted in Fig. 2. The anomalies can be separated into two groups for temperatures:

i) $T < 13$ K, characterized by a magnetization step $\Delta M$ across the transition and a field-hysteresis $\Delta H = (H_{c\uparrow} - H_{c\downarrow})$, where $H_{c\uparrow}$ ($H_{c\downarrow}$) represents the critical field of the transition when sweeping the field up (down). $\Delta H$ decreases with increasing temperature and vanishes between 12 and 13 K as seen in the lower right inset of Fig. 2. As a critical field of this transition, $H_c$, we take the average of $H_{c\uparrow}$ and $H_{c\downarrow}$ values. When approaching the FOMPT from low fields the $M(H)$ dependence is almost linear up to at least 80% $H_c$ and then follows a weak upturn and a steep flank with an inflection point at $H_c$. The $M(H)$ curve for $H > H_c$ is concave.

ii) 13 K $< T <$ $T_N$, involving a progressively pronouncing upturn of the $M(H)$ curve terminated by a cusp at a field which we consider as $H_c$. The transition has no hysteresis. $H_c$ decreases with increasing temperature to become 0 T at $T_N$. The upturn and the cusp become simultaneously less pronounced with increasing temperature and vanish at 16.5 K. The $M(H)$ curves are convex in a certain field interval above $H_c$.

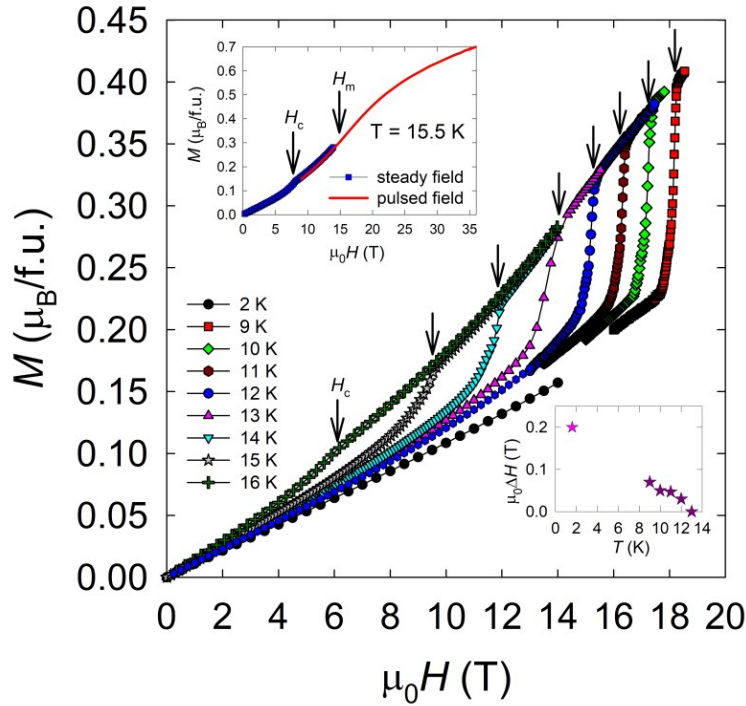

Fig. 2. Selected magnetization isotherms, which are characteristic of anomalies reflecting field-induced phase transitions (metamagnetic transitions), measured in static magnetic fields applied along the $b$ axis. The vertical arrows mark the positions of $H_c$. For sake of clarity only selected isotherms are displayed. Left inset: Comparison of 15.5-K data from steady and pulsed magnetic

fields. Right inset: Temperature dependence of hysteresis of MT (Magenta point at temperature 1.6 K was taken from heat capacity data).

Selected magnetization isotherms measured at various temperatures in pulsed fields up to 35 T are shown in Fig. 3a. Each magnetization curve from temperatures below 13 K exhibits a pronounced magnetization step with a large hysteresis of several T. The $M(H)$ curves at temperatures above 13 K have a broad S shape with an inflection point at a field which we mark $H_m$. In the left inset of Fig. 2 one can see how the steady-field and pulsed-field data at higher temperatures relate. A particular $H_m$ value can be well determined as a field of the $dM/dH$ maximum (see Fig. 3b). The S shape of the $M(H)$ curves persists to temperatures well above $T_N$ but becomes gradually smeared out with increasing temperature and vanishes above 30 K. At higher temperatures, the magnetization data follow a flat concave function of magnetic field over the entire field range as expected for a normal paramagnet.

No clear signature of a cusp at $H_c$ (observed in static fields) has been detected in the pulsed-field data collected at temperatures between 13 K and $T_N$. The sensitivity of the pulsed-field experiment and the following data acquisition are apparently not sufficient to distinguish weak magnetization anomalies in low fields, which can be detected in static fields.

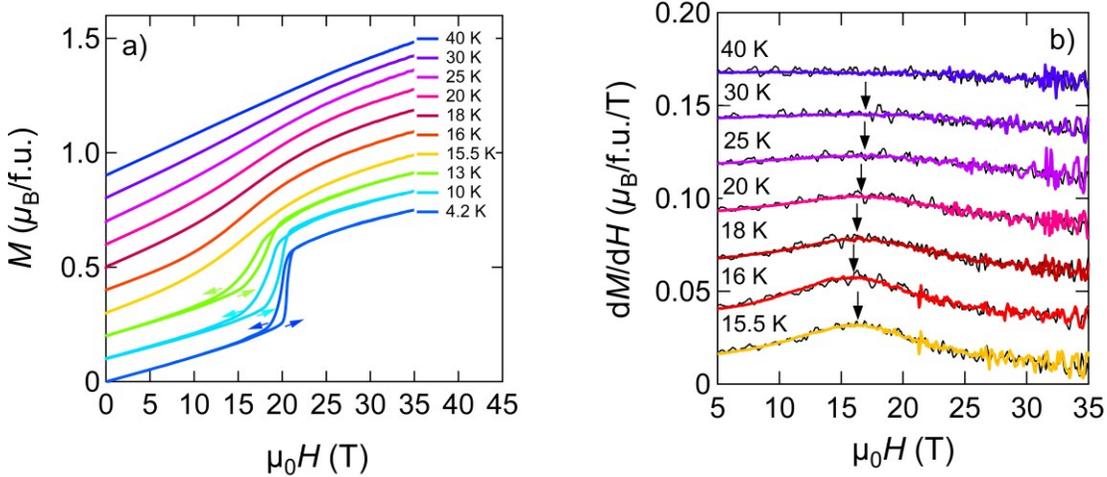

Fig. 3. a) Selected magnetization isotherms measured in pulsed magnetic fields applied along the $b$ axis. b) Derivatives of magnetization isotherms measured in pulsed magnetic fields applied along the $b$ axis at temperatures above 13 K. The arrows mark the positions of $H_m$.

B. Magnetocaloric effect – MCE

The pulsed-magnetic-field heat-capacity setup at IMGSL offers a possibility to measure MCE as an important tool for determination of the character of a magnetic phase transition[15, 16, 20, 21]. A first-order phase transition, where dissipative mechanisms are present, shows characteristic asymmetry due to the release of the heat in both directions of the magnetic field sweep[16]. MCE data also serve as an important base for assessment of reliability of measurements in pulsed magnetic fields.

The MCE in UIrGe was measured below $T_N$ with the initial (zero field) temperature $T_{init}$ of 1.8 and 14.5 K, respectively and above $T_N$ at the initial temperature of 17 K (Fig. 4). For $T_{init}$ = 1.8 K with the magnetic field sweeping up, the sample temperature reached 2.5 K when the field approached $H_c$ and then the sample temperature suddenly dropped by $\delta T \sim$ -1.7 K at MT. On the other hand, a sudden sample warm up by $\delta T \sim +3.5$ K has been recorded when the magnetic field was sweeping down through MT (Fig. 4), i.e. MCE is considerably asymmetric with respect to the direction of magnetic field sweep. For $T_{init} = 14.5$ K, MCE shows a shallow valley with a minimum near the phase-transition line in the magnetic phase diagram and almost symmetric $\delta T = \pm 1.1\text{-}1.2$ K. In the case of $T_{init} = 17$ K the valley is very shallow (< 1 K), although still recognizable, and the MCE is also symmetric. The minimum temperature appears near to $H_m$.

The nonsymmetric response of MCE to MT in UIrGe for $T_{init} = 1.8$ K indicates that this transition is at low temperatures of the first order type. On the other hand such MCE hampers the measurements of magnetization isotherms at low temperatures when the sample is placed only in He exchange gas. In this case the desired isothermal conditions are not accomplished during the field pulse due to eddy currents heating and the asymmetric MCE in the neighborhood of FOMPT. This is well documented by comparison of the 10-K and 13-K $M(H)$ curves obtained in static fields (Fig. 2) and in pulsed fields (Fig. 3a). The magnetization step due to MT in the first case is sharp and symmetric, and shows a hysteresis of less than 0.1 T in contrast to a very broad nonsymmetric anomaly with a hysteresis of several T. The 4.2-K curve measured with the sample immersed in He liquid is considerably sharper but shows again very large and non-symmetric hysteresis. For the above reasons, we have not included the pulsed-field magnetization data at temperatures ≤ 13 K to phase diagram and worked only with the magnetization data obtained in static fields and heat-capacity data recorded in quasistatic fields at 1.6 K. On the other hand, for the magnetization data obtained in pulsed fields at higher temperatures ≥ 15.5 K, where a negligible and symmetric MCE has little influence on sample temperature, we conveniently used especially the higher-field data for determination of $H_m$.

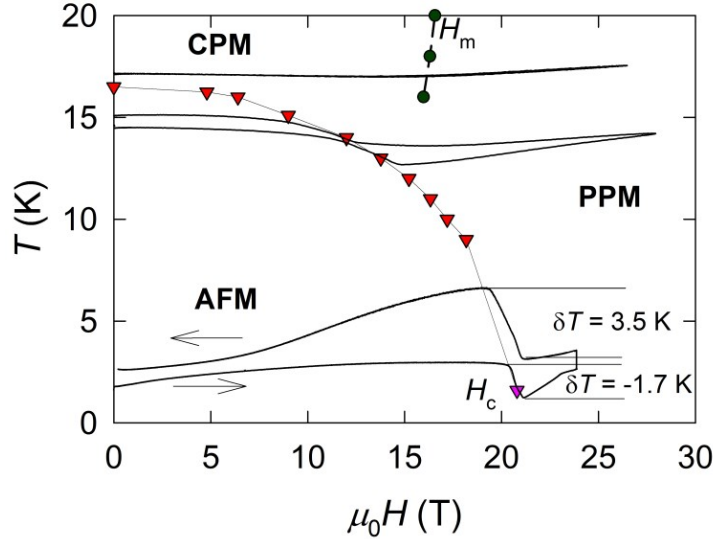

Fig. 4. Results of selected MCE scans with different initial temperatures $T_{init}$. The full lines are the curves representing sample temperature during the scans. Red tringles represent the critical field $H_c$ of MT, $H_m$ inflection point of the magnetization S-shape, magenta point at temperature 1.6 K was taken from heat capacity data. Arrows show the direction of field sweeps.

C. Heat capacity in high fields

The magnetic-field dependence of heat capacity $C_p/T(H)$ at $T = 1.6$ K shown in Fig. 5 exhibits a large jump at 20.8 T, which is apparently the critical field $H_c$ of the metamagnetic transition. The data points from the close neighborhood of the transition, which are shown in the inset of Fig. 5, form two spikes, one for field sweep up and one for sweep down. These data are an artefact of the data acquisition and evaluation method applied in the experiment and have no real meaning in terms of heat capacity of the measured material. Nevertheless, the field difference of ~0.2 T between the spikes can be considered as a reasonable estimate of the hysteresis of the MT in UIrGe at 1.6 K. When analyzing the data from the experiment we have revealed a subtle temperature instability (1.6 ± 0.2 K) which is the most probable cause of the extrinsic broad bump of $C_p/T(H)$ in fields below MT.

The $C_p/T(H)$ jump at $H_c$ at 1.6 K may be understood as a sudden enhancement of Sommerfeld coefficient $\gamma$ of UIrGe from the zero-field value (= 16 mJ/mol K$^2$) to ≈ 90 mJ/mol K$^2$ in $H > H_c$, which is accompanying the MT. The temperature scan of the heat capacity data (not shown) in the flat-top pulsed field 24 T ($> H_c$) confirmed the value ≈ 90 mJ/mol K$^2$.

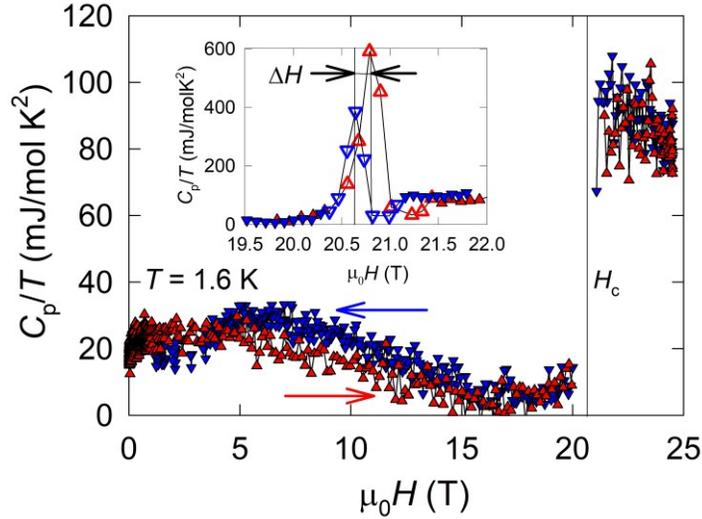

Fig. 5. Magnetic-field scans of the heat capacity of UIrGe at 1.6 K up to 25 T applied along the $b$ axis. MT appears at 20.8 T at $H_c$ as a sudden increase of $C_p/T$. Only intrinsic data are plotted. In inset: data in the vicinity of $H_c$ drown by empty triangles, which are due to the artefact of the experiment.

## IV. DISCUSSION

The collection of characteristic parameters of magnetism in UIrGe, $T_N$, $H_c$, and $T_{max}$, $H_m$ and $T_{dev}$ determined by aforementioned measurements enabled us to construct a magnetic phase diagram in the $T-H$ plane for magnetic fields along the $b$ axis, which is displayed in Fig. 6. The $H_c$ values at various temperatures determined from $M(H)$ isotherms and the $T_N$ values coming from $C_p(T)$[7, 8] dependences in various fields determine the phase-transition line between the AFM and PM phases. The metamagnetic transitions at $T < 13$ K and corresponding fields $\mu_0 H > 15$ T exhibit a field hysteresis, a hallmark of FOMPT, whereas the transitions at higher temperatures $T > 13$ K and lower magnetic fields $\mu_0 H < 15$ T show no trace of hysteresis as expected for SOMPT. The conclusion about FOMPTs at temperatures below 13 K is corroborated by observing the asymmetric signal of the MCE observed in the vicinity of the transition (Fig. 4).

It is worth noting that the temperature dependences of heat capacity measured in various $b$-axis fields up to 17.5 T[7, 8] corroborate our conclusion about the order of the AFM↔PM transition. The usual attribute of SOMPT, a $\lambda$-anomaly in $Cp/T$ vs $T$, is observed in fields up to 14 T ($T_N > 13$ K) whereas the symmetric peaks that show up at lower temperatures and higher fields correspond to FOMPT.

Besides the hysteresis, also some other features of $M(H)$ dependences differentiate between the FOMPT and SOMPT segments of the magnetic phase diagram (see Figs. 2 and 3). In the case of FOMPT ($T \leq 12$K) the $M(H)$ dependences are almost linear up to high fields and show a weak upturn only when approaching MT followed by a steep flank with inflection point at $H_c$ whereas in the SOMPT segment we observe $M(H)$ curves with an upturn progressively developing already from low fields and terminated by a kink at $H_c$.

One possible explanation of the $M(H)$ upturn may be considered an effect caused by field-induced flips of individual spins from the AFM arrangement to the field direction. The number of the slips increases with increasing field toward SOMPT at $H_c$. In the case of FOMPT, the field-induced spin flips in fields well below $H_c$ are very rare, almost negligible.

A possibility that these different responses of UIrGe to low applied magnetic fields ($< H_c$) may be connected with two different AFM phases needs to be thoroughly tested by neutron scattering or µSR in magnetic fields. The $M(H)$ dependence in fields above $H_c$ of SOMPT is convex up to $H_m$ whereas it is always concave in the FOMPT case.

The appearance of both the FOMPT and SOMPT segments in the magnetic phase diagram is expected in Ising antiferromagnets with competing interactions and the point separating the two segments of the AFM ↔ PM phase-transition line has been interpreted as a tricritical point[22]. Experimental studies concerning tricriticality in antiferromagnets are quite rarely reported in literature although understanding of these phenomena has fundamental importance. TCPs have been reported in several uranium antiferromagnets exhibiting Ising-like behavior caused by very strong magnetocrystalline anisotropy[23-27]. We mark the tentatively considered TCP in the magnetic phase diagram in Fig. 6 by a yellow circle with coordinates [$T_{tc}$, $H_{tc}$]. It is worth mentioning that TCP is also identical with the inflection point of the wrapping curve of the series of magnetization isotherms displayed in Fig. 2.

The magnetization isotherms of a material in a normal paramagnetic state follow some modification of the Brillouin function, which is always concave. In the low-temperature range, $T < 13$ K the $M(H)$ curve for $H > H_c$ is concave (Fig. 2) saturating with further increasing $H$ as expected for a paramagnet with magnetic moments further gradually polarized by the applied magnetic field. We call the respective $T$-$H$ space section ($T < 13$ K, $H > H_c$) the polarized paramagnet regime (PPM).

From Fig. 3a it is evident that the magnetization isotherms of UIrGe measured at temperatures above 13 K in fields $H > H_c$ are convex up to a characteristic field $H_m$ above which they become concave and gradually saturate with further increasing field. Consequently, the $M(H)$ curves in fields above $H_c$ have a typical broad S shape which becomes gradually smeared with increasing temperature and vanishes at temperatures above 30 K. $H_m$ is determined as the inflection point of the S shape. Both the broad S-shape anomalies on magnetization isotherms (Fig. 3) and broad maxima on temperature dependence of magnetic susceptibility (Fig. 1) are characteristic of materials near a magnetic instability connected with metamagnetism of various microscopic origins. UIrGe in $H//b$ is in our opinion one such cases in which MT is the transition between the low-field AFM phase and the high-field paramagnetic phase. In the experiment we observe that both, $T_{max}$ and $T_N$ decrease with increasing field. $T_{max}$ simultaneously approaches $T_N$ to become equal $T_{max} \approx T_N$, at the tentatively considered TCP that separates the high-temperature segment of the SOMPTs and the low-temperature segment of the FOMPTs. Interestingly, also the line representing the temperature-induced evolution of $H_m$ is bound for this point in the magnetic phase diagram.

To explain this unusual magnetization/susceptibility behavior we suggest the following tentative scenario: $\chi(T)$ in a normal paramagnetic state follows a MCW law. UIrGe indeed behaves like that at high temperatures above ~ 50 K. The best fit of $\chi_b(T)$ data above 55 K to a MCW law leads to the values of effective magnetic moment $\mu_{eff}$ = 2.52 $\mu_B$/f.u., Weiss temperature $\Theta_p$ = -34 K and temperature independent susceptibility $\chi_0 = 2.5 \times 10^{-9}$ m$^3$/mol[4].

When decreasing temperature below $T_{dev}$, the $\chi(T)$ data progressively deviate downwards from the MCW law, show a broad maximum at $T_{max}$ and suddenly drop at $T_N$ due to onset of AFM ordering. The behavior between $T_{dev}$ and $T_N$ is apparently caused by some mechanism which is reducing the susceptibility. Application of magnetic field leads to gradual enhancement of $\chi_b(T)$ values for $T_N < T < T_{dev}$ and consequently reduced deviation from MCW behavior. The evolution of susceptibility correlates with the convex shape of magnetization curves between $H_c$ and $H_m$. An effect of (dynamic) AFM correlations between magnetic moments (or static short-range AFM order) in the paramagnetic state may be tentatively considered as a possible explanation. The AFM correlations (or static short-range AFM order) may cause some originally paramagnetic moments couple antiferromagnetically with their counterparts. This process in fact may lead to a reduction of susceptibility due to reduction of the number of paramagnetic moments by the number of AFM coupled moments. In case of AFM correlations (short-range AFM order) this process would be dynamic (static). The AFM correlations (short-range AFM order) are progressively enhanced with decreasing temperature from $T_{dev}$ to $T_N$ and suppressed with increasing magnetic field between $H_c$ and $H_m$. The phase space bordered by the $T_N$-$T_N(H_{tc})$ phase transition line, $H_m(T)$ and $T_{dev}(H)$ crossover lines we tentatively call the correlated paramagnet (CPM) regime.

We are fully aware that our scenario for UIrGe supported by presently available data collection which has been delivered by only macroscopic measurements, is rather a speculation. Nevertheless, we hope that presenting it brings new motivation for thorough investigations of UIrGe by microscopic methods. Conclusive information about AFM correlations (or short-range AFM order) in the CPM regime and the CPM↔PM crossover is provided by inelastic neutron scattering like in the case of the heavy fermion antiferromagnet $U_2Zn_{17}$[28]. The disappearance of characteristic excitations in neutron spectrum can be taken as a reliable parameter of CPM↔PPM crossover especially when investigating it in applied magnetic fields. The short-range AFM order can be studied e.g. by low-angle magnetic neutron scattering.

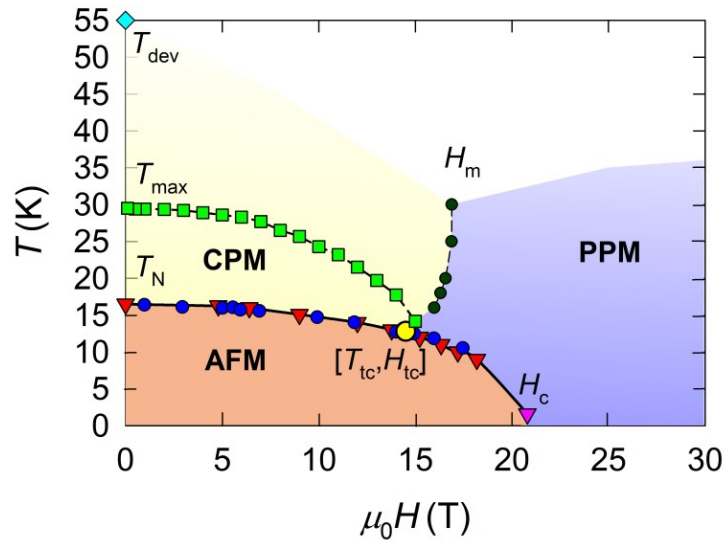

Fig. 6. The UIrGe $T$-$H$ phase diagram for magnetic fields applied along the $b$ axis. The red triangles represent the $H_c$ values of MTs determined by measurements of $M(H)$ isotherms in static magnetic fields. The magenta triangle corresponds to $H_c$ of the MT detected at 1.6 K by the heat capacity measurement in pulsed fields. The average of $H_{c\uparrow}$ and $H_{c\downarrow}$ values is displayed as $H_c$ in case of FOMPTs. The dark blue circles represent the $T_N$ determined from $C_p(T)$ data in static magnetic fields reported in Refs[7, 8]. The dark green circles correspond to $H_m$ values from pulsed-field $M(H)$ data. The green squares represent the $T_{max}$ values, temperatures of maxima of $\chi(T)$ dependences measured in static magnetic fields. The yellow circle with coordinates [$T_{tc}$,$H_{tc}$] is the point between the low temperature region with FOMPT and high temperature region SOMPT. The light blue diamond corresponds to the temperature $T_{dev}$, below which the $\chi(T)$ data deviate from the high-temperature data (> 55 K) fit to a MCW law. Rigorous determination of $T_{dev}$ is quite difficult. Therefore we shown only the point in the low-field limit. The upper edge of the yellow shading tan be considered as a roughly estimated field dependence of $T_{dev}$.

The unique measurement of the heat capacity in pulsed fields (Fig. 5) gave us direct evidence about the change of the value of Sommerfeld coefficient $\gamma$ due to MT. The $C_p/T$ value at 1.6 K, which can be in our case a reasonable estimate of $\gamma$, undergoes a sudden jump in the vicinity of $H_c$ when magnetic ordering is destroyed by magnetic field. The $C_p/T$ is stabilized at ~90 mJ/molK$^2$ above $H_c$, which corresponds to the extrapolated $C_p/T$ (0 T) value from temperatures above $T_N$ (0 T)[4]. We tentatively attribute this result to Fermi surface (FS) reconstruction although we are aware that also some other mechanism, e.g. field-induced change of magnetic fluctuations, can cause the change of $\gamma$. As a supporting argument for our scenario we would like to add the consideration that UIrGe as a U intermetallic compound could have the 5$f$-electron (carrying magnetic moments) states present at the Fermi surface. Any change of magnetic periodicity (AFM transition at $T_N$, metamagnetic transition from AFM to paramagnetic state) should be connected with FS reconstruction. In any case, additional measurements (e.g. magnetoresistance, Hall, Seebeck, and de Haas-van Alphen effect, angle-resolved photoemission spectroscopy) associated with relevant electron-structure calculations are needed to test validity of our suggestion.

The recovery of the enhanced $\gamma_{UIrGe}$ was deduced also by an empirical Kadowaki-Wood relation[29] above the critical hydrostatic pressure $p_c \approx 12$ GPa, where the AFM phase vanishes and AFM gap closes[14]. The estimated value $\gamma_{15GPa} = 80$ mJ/molK$^2$ is close to the extrapolated ambient-pressure $C_p/T$ value from the paramagnetic range.

## V. Conclusions

We have studied various aspects of magnetism in the $T$-$H$ phase space of the antiferromagnet UIrGe in steady and pulsed magnetic fields applied along the orthorhombic $b$ axis by measurements of magnetization, heat capacity and MCE on a UIrGe single crystal. The obtained results have been used to draw the $T$-$H$ magnetic phase diagram.

UIrGe is antiferromagnetic below $T_N$ = 16.5 K; $T_N$ decreases with increasing magnetic field. The magnetic field-induced transition from the AFM to the paramagnetic state change its character at 13 K. The MT is a first-order phase transition at low temperatures with a typical large step of magnetization at $H_c$ and magnetic field hysteresis. In magnetic fields higher than $H_c$, the magnetic moments of the paramagnet are polarized by the magnetic field - the system is in a PPM regime and magnetization gradually saturates in higher fields as a result of magnetic-field influence on the band structure.

At 13 K the character of the field-induced transition from the antiferromagnetic to paramagnetic phase changes considerably. In the temperature interval 13 K < $T$ < $T_N$, it is a SOMPT as manifested by a pronounced upturn of the *M(H)* curve terminated by a kink at $H_c$. The broad S-shaped magnetization is detected in magnetic fields above $H_c$ in the field $H_m$ of the crossover between CPM and PPM. The $H_m$ crossover line in the *T-H* space is found to be bound to the point at [~ 13 K, ~ 15 T] where the type of MT changes from a FOMPT to a SOMPT. In low magnetic fields the CPM is demarcated by the temperature of the onset of deviation of $\chi(T)$ from the high temperature around which UIrGe passes a crossover between CPM and the normal PM regime existing at higher temperatures.

The FOMPT is accompanied by a sudden enhancement of Sommerfeld coefficient demonstrated by a sudden increase of the $C_p/T$ value at 1.6 K from the ground-state value of 16 mJ/mol K$^2$ to ~ 90 mJ/mol K$^2$, which indicates a possible Fermi-surface reconstruction. The direct observation of the band structure changes induced by magnetic phase transition by ARPES, de Haas-van Alphen, Hall, and Seebeck effect measurements will be the subject of further investigation of UIrGe as well as the neutron spectroscopy investigation of the AF correlations in the CPM regime.

**Acknowledgements**

This work was supported by JSPS KAKENHI Grant Numbers 15H05884 (J-Physics) and 16K05463. The authors are indebted to Mohsen M. Abd-Elmeguid for fruitful discussion of the experimental data and Ross H. Colman for a critical reading of the manuscript and language corrections. Magnetization measurements in static magnetic fields up to 18.5 T were performed in the Materials Growth and Measurement Laboratory MGML (see: http://mgml.eu).